\documentclass{aa}

\newcommand* {\Msun} {\mbox{M$_\odot$}}
\usepackage{graphicx}
\usepackage{txfonts}
\usepackage{hyperref}
\hypersetup{
     colorlinks=true,
     linkcolor=blue,
     filecolor=blue,
     citecolor = blue,      
     urlcolor=blue,
     }
\begin{document}

   \title{Pattern speed evolution of barred galaxies in TNG50}

   \author{Marcin Semczuk
          \inst{1,2,3}\fnmsep\thanks{email: marcin.semczuk@yahoo.com}
          \and
            Walter Dehnen\inst{4}
            \and 
            Ralph Sch{\"o}nrich\inst{5}
            \and
            E. Athanassoula\inst{6}
          }

   \institute{Departament de Física Qu\`antica i Astrof\'isica (FQA), Universitat de Barcelona (UB),  c. Mart\'i i Franqu\`es, 1, 08028 Barcelona, Spain
         \and
Institut de Ci\`encies del Cosmos (ICCUB), Universitat de Barcelona (UB), c. Mart\'i i Franqu\`es, 1, 08028 Barcelona, Spain
        \and
Institut d'Estudis Espacials de Catalunya (IEEC), c. Gran Capit\`a, 2-4, 08034 Barcelona, Spain
        \and
Astronomisches Rechen-Institut, Zentrum f{\"u}r Astronomie der Universit{\"a}t Heidelberg, M{\"o}nchhofstra\ss{}e 12-14, 69120, Heidelberg, Germany
        \and
Mullard Space Science Laboratory, University College London, Holmbury St.~Mary, Dorking, Surrey, RH5 6NT, UK
        \and
Aix Marseille Univ, CNRS, CNES, LAM, Marseille, France
             }

   \date{submitted to A\&A}


\abstract
{Galactic bars are found in the majority of disc galaxies. They rotate nearly rigidly with an angular frequency called pattern speed. Previous idealised simulations have shown that bar pattern speed decreases with time due to dynamical friction exerted by the dark matter halo, while cold gas can reduce or even reverse this trend.}
{We want to understand how different galaxy properties affect the evolution of the bar pattern speed in more realistic situations, including ongoing star formation, mass infall, AGN feedback and galaxy interactions.}
{We used the high-resolution run TNG50-1 of the magnetohydrodynamical cosmological simulations suite IllustrisTNG to trace the pattern speed of simulated bars and see how it depends on various galaxy properties.}
{Simulated bars with initially high pattern speed and a subsequent rapid slowdown are more likely found in more massive galaxies.
Lower mass galaxies, on the other hand, preferentially host bars that start at relatively low pattern speeds and retain the same value until the end of the simulation. More massive barred galaxies are also more affected by the AGN feedback, which removes (or heats up) the cold gas that could have prevented the slowdown.  
}
{We find that bars grow and strengthen with slowdown, in agreement with higher resolution simulations. We find that strong correlations between the bar slowdown rate and galaxy mass weaken considerably when we use dimensionless measures to quantify the slowdown. In TNG50, the AGN feedback prescription amplifies the mass dependence. Turned around, this provides an interesting statistic to constrain subgrid physics by bar growth/slowing.}

   \keywords{Galaxies: bulges --
                Galaxies: evolution --
                Galaxies: structure --
                Galaxies: kinematics and dynamics --
                Galaxies: spiral
               }

   \maketitle
%

\section{Introduction}
Bars are ubiquitous in spiral galaxies in the local universe with roughly two-thirds of them hosting a bar in their inner disc \citep{EskridgeEtal2000, MenendezDelmestreEtAl2007, ShethEtAl2008, Masters2011, Cheung2013, Erwin2018}. Recent discoveries by the JWST show that bars are also frequent at higher redshifts \citep{Guo2023,Costantin2023,LeConte2024}.

The rotation rate or `pattern speed' $\Omega_\mathrm{p}$ of bars was initially believed to stay constant with time, but numerical simulations of isolated galaxies have shown that $\Omega_\mathrm{p}$ decreases owing to the dynamical friction of the bar against the dark matter halo \citep{Sellwood1980, Weinberg1985, LittleCarlberg1991, DebattistaSellwood1998, Athanassoula2003}. Cold gas, on the other hand, loses its angular momentum to the bar as it is driven into the nuclear region, which reduces or may even reverse the bar slowdown and tends to diminish the bar strength \citep{Friedli1993, VV2009, VV2010, Athanassoula2013, Athanassoula2014, Beane2023}.

A dimensionless measure for the bar rotation rate is the ratio $\mathcal{R}=R_\mathrm{CR}/L_\mathrm{bar}$ of the corotation radius $R_\mathrm{CR}$ (where the circular frequency of stars rotating in the disc equals $\Omega_{\mathrm{p}}$) and the bar length $L_\mathrm{bar}$. Theoretical arguments based on the extent of bar-supporting orbits prescribe that $\mathcal{R}>1$ \citep{Contopoulos1980}. Bars with $\mathcal{R}\lesssim 1.4$ are conventionally classified `fast' and those with $\mathcal{R}>1.4$ `slow' \citep{DebattistaSellwood2000}. While observational estimates of $\mathcal{R}$ are plagued by some difficulties and may in some cases lead to findings of `ultra-fast' bars ($\mathcal{R}<1$, \citealt{Cuomo2021}), most observed bars appear to be fast (\citealt{Corsini2011, Aguerri2015, Guo2019}). This seems to be at odds with emerging cosmological simulations which produce slow 
\citep{Algorry2017, Peschken2019, Roshan2021} or short bars \citep{Frankel2022}. Findings of \cite{Fragkoudi2021} suggest that this tension lessens with increasing resolution of the simulated barred galaxies.  

In this study, we trace the pattern speeds of a sample of simulated barred galaxies from the TNG50-1 run of the IllustrisTNG suite of simulations \citep{Pillepich2019, Nelson2019b, Nelson2019}. We aim to identify factors which affect the time evolution of $\Omega_{\mathrm{p}}$ and to relate them to the knowledge learned from isolated idealised $N$-body experiments. One goal of this work is to understand how the astrophysics implemented in TNG50 via sub-grid models affects the pattern speeds of bars and their time evolution. 

The paper is structured as follows: Section 2 briefly introduces the simulations and the sample of simulated barred galaxies that we considered in our analysis. Section 3 details how we measure $\Omega_\mathrm{p}$. In section 4 we correlate averaged properties of bar slowdown and host galaxies and identify the most relevant ones. Section 5 describes the relation between bars simulated in TNG50 and AGN feedback that we found to occur in the simulation. In section 6 we discuss our findings in a broader context and summarize them.     

\section{Simulations}
IllustrisTNG is a suite of magnetohydrodynamical cosmological simulations that tracks the formation and evolution of galaxies. Besides gravity and magnetohydrodynamics it also numerically approximates via so-called sub-grid models many other physical processes that are relevant to galaxy evolution, such as star formation, growth of Supermassive Black Holes (SMBHs), and feedback from both AGN and stars. The run TNG50-1 has the highest resolution with typical stellar particle mass of $8.5\times10^4\,\Msun$ and a gravitational softening length of 288\,pc. Different aspects of bars in TNG50-1 were previously studied by  \cite{RG22, Frankel2022, IV2022, Zana2022, Ansar2023, Lopez2024, RG24}.

\subsection{Sample selection}
First, we use a list of $105$ galaxies with bars at $z = 0$ compiled by \cite{RG22}. We measure the azimuthal Fourier transform $\hat{\Sigma}_m$ of the surface density of these galaxies at snapshot 86 (which is $z=0.17$, i.e. $\sim2.1$ Gyr before the end of the simulation), obtaining the relative $m=2$ amplitude $A_2\equiv|\hat{\Sigma}_2|/\hat{\Sigma}_0$. To ensure sufficient signal, we cut the sample to those 79 simulated galaxies for which the maximum $A_2>0.2$ and for which the number of star particles inside the radius of this maximum exceeds $5\times10^4$. This excludes objects in the original sample which form a bar only later, such that their bar history is too short to meaningfully study bar evolution.

\begin{figure}
\centering
\includegraphics[width=\columnwidth]{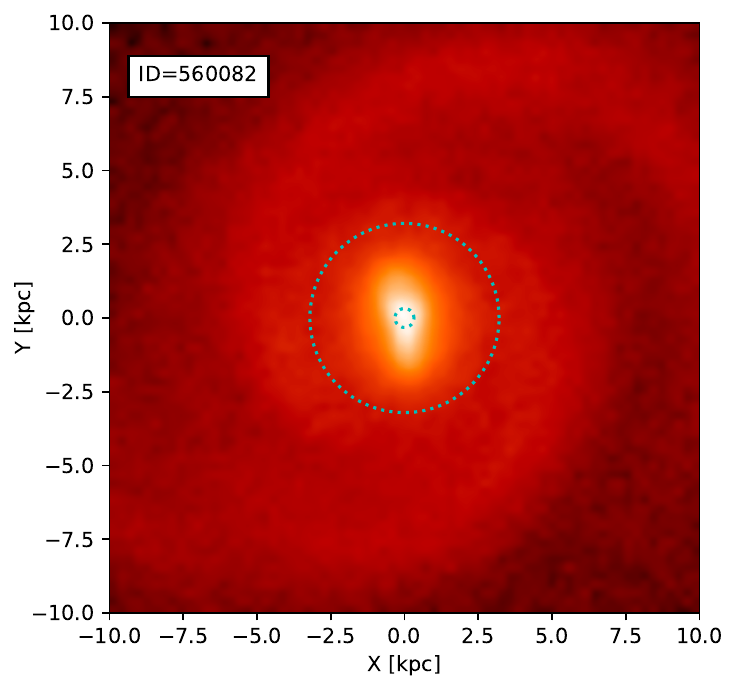}
\caption{An example barred galaxy at $z=0$ with marked boundaries of the bar region derived as outlined in section 3. The maximum of $A_2$ amplitude for this snapshot is 0.38.}
          \label{bar_reg}%
\end{figure}

After this initial selection, several simulated galaxies still have several snapshots where our method did not work. Visual inspection revealed that the measurements were affected by various factors substantially disturbing the bar shape. These factors were strong interactions with external galaxies passing through(minor mergers), very strong $m=2$ spiral arms, and double bars (see \citealt{Semczuk2024}). Additionally, a few other cases have very weak, round bars that marginally pass $A_2>0.2$ in some snapshots, but are too weak in others to allow tracing the continuous evolution of the pattern speed. After rejecting these cases we ended up with a final sample of 62 simulated galaxies with a stellar mass covering the range of $10^{10}\text{-}10^{11.2}\,\mathrm{M_\odot}$ at $z=0$.

\section{Measurements of the bar pattern speeds}
To track the evolution of the bar pattern speeds $\Omega_{\mathrm{p}}$ we employ the program \texttt{patternSpeed.py} \citep{Dehnen2023}, version 0.5.3. The method used to identify the bar region (from which $\Omega_{\mathrm{p}}$ is measured) is somewhat different from that described by \citeauthor{Dehnen2023}. In particular, we measure for a range of overlapping annuli the relative azimuthal Fourier amplitudes, $A_m\equiv|\hat{\Sigma}_m|/\Sigma_0$, of the surface density and define for each annulus the bar strength as 
\begin{align}
    S = \left[A_2^2+A_4^2+A_6^2\right]^{1/2} - \left[A_1^2+A_3^2+A_5^2\right]^{1/2}.
\end{align}
If $S>0.1$ at any radius, the bar region is identified as the range of annuli for which $S>0.025$ and the Fourier phases $\psi_{m=2}$ are within a 15$^\circ$ interval.

\begin{figure*}
\centering
\includegraphics[width=18.4cm]{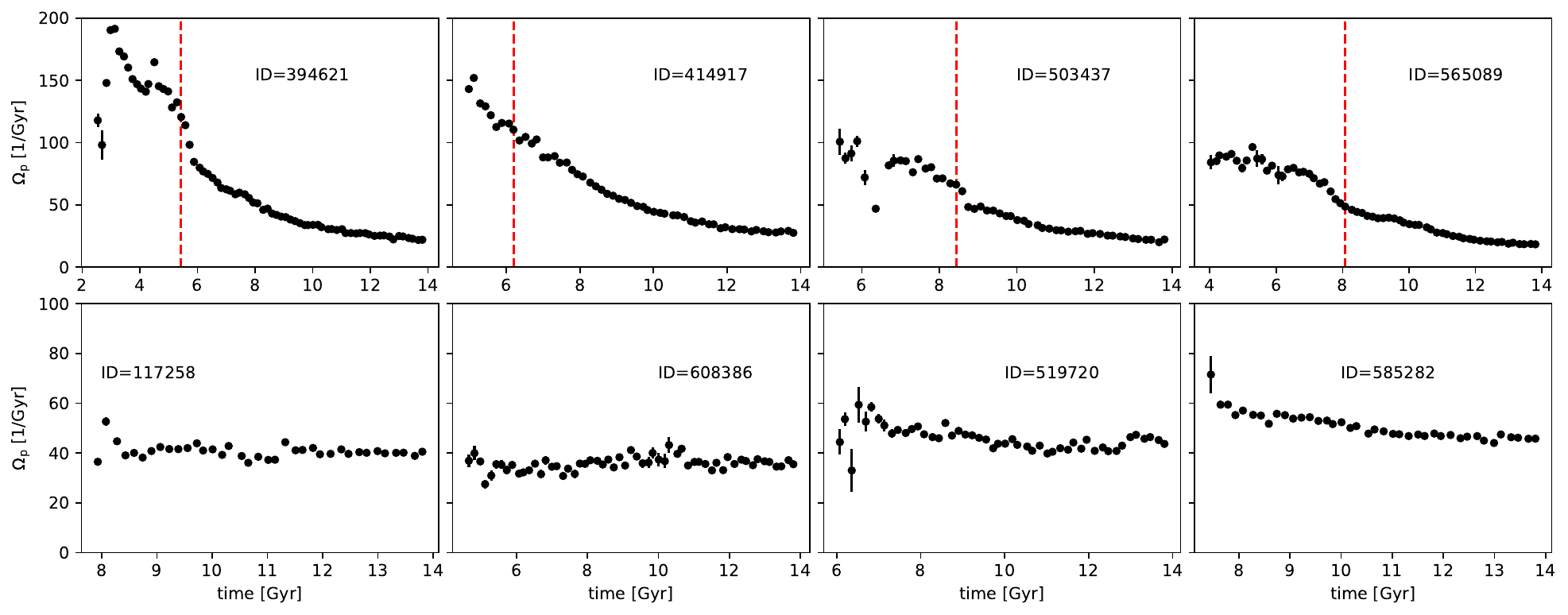}
\caption{Eight examples of bar-pattern-speed evolution (the IDs given are for $z=0$) with significant slowdown (top) or little evolution (bottom). The time axis begins with the first snapshot where a bar is detected. Dashed vertical red lines indicate the time $t_\mathrm{AGN}$ when kinetic feedback from the AGN reaches a threshold that coincides the appearance of a hole in the simulated gas discs.
}
\label{omegas}%
\end{figure*}

\begin{figure*}
\centering
\includegraphics[width=18cm]{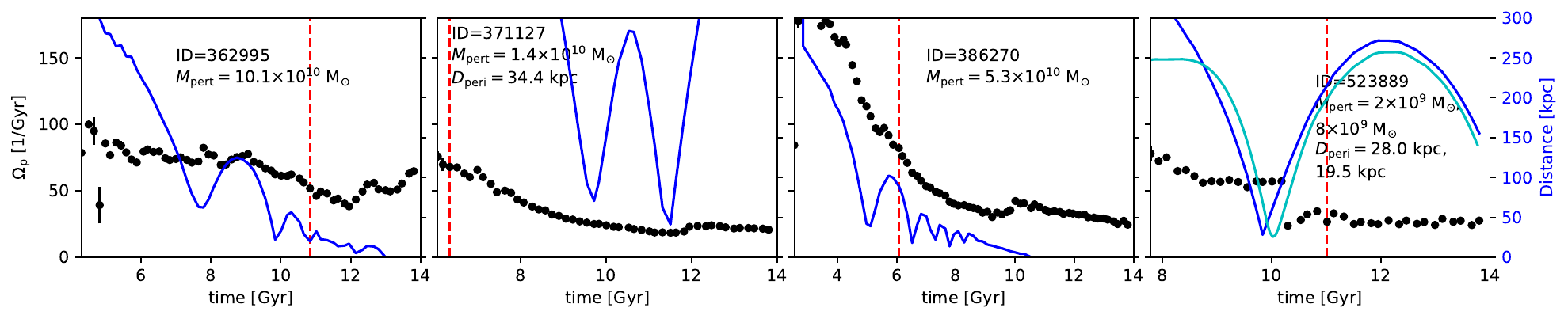}
\caption{As Fig.~\ref{omegas} for four cases with somewhat irregular pattern-speed evolution. Blue and cyan lines show the relative distance between the host barred galaxies and an impacting interacting galaxy, whose total maximal masses $M_{\mathrm{pert}}$ prior to the interaction and distance of closest approach are indicated.
}
\label{jumps}%
\end{figure*}

Figures~\ref{omegas} and~\ref{jumps} show typical pattern-speed evolutions for bars from the considered sample (dots). Also shown (vertical red dashed lines) is the time $t_\mathrm{AGN}$, at which the cumulative amount of kinetic AGN feedback energy injected into surrounding gas reaches the threshold of $10^{15}$ $\mathrm{M}_\odot/h (\mathrm{ckpc}/h)^2/(0.978 \mathrm{Gyr}/h)^2$. We find that this coincides well with the carving out of the gas hole in the simulated discs, caused by the kinetic feedback (in conjunction with the fact that the cold gas phase is not realistically modelled at the resolution of these simulations). A larger bar slowdown occurs preferentially after $t_\mathrm{AGN}$, as we discuss in Sections 4 and 5. In the top row of Fig.~\ref{omegas} we show four cases where the bar starts rotating at high absolute values of $\Omega_\mathrm{p}\geq100\,$Gyr$^{-1}$ and then drastically slows down to $\sim50\,$Gyr$^{-1}$. Interestingly \cite{Guo2019} find a few observed galaxies with $\Omega_\mathrm{p}\sin i >80\,$Gyr$^{-1}$ which would match the values of these simulated bars in the rapid slowdown phase. In the bottom panel bars exhibit a different behaviour: their $\Omega_\mathrm{p}\simeq40\text{-}60\,$Gyr$^{-1}$ remains approximately constant throughout the evolution.

Four examples with more irregular behaviour are shown in 
Fig.~\ref{jumps}, where we also plot (blue) the time evolution of the distances to the a interacting or merging companion galaxies. The first and third cases experience a minor merger (that ends at $\sim13$ and $\sim10$ Gyr respectively), while the second and fourth are subject to fly-bys. The result of these interactions may vary from a wiggle pattern in case one, through small jumps in cases two and three to a step-like jump in case four. We note that case four is the most extreme jump within our sample. These irregularities increase the noise when one attempts to find the main determinants for the secular evolution of bar pattern speed.

\begin{figure*}
\centering
\includegraphics[width=17.5cm]{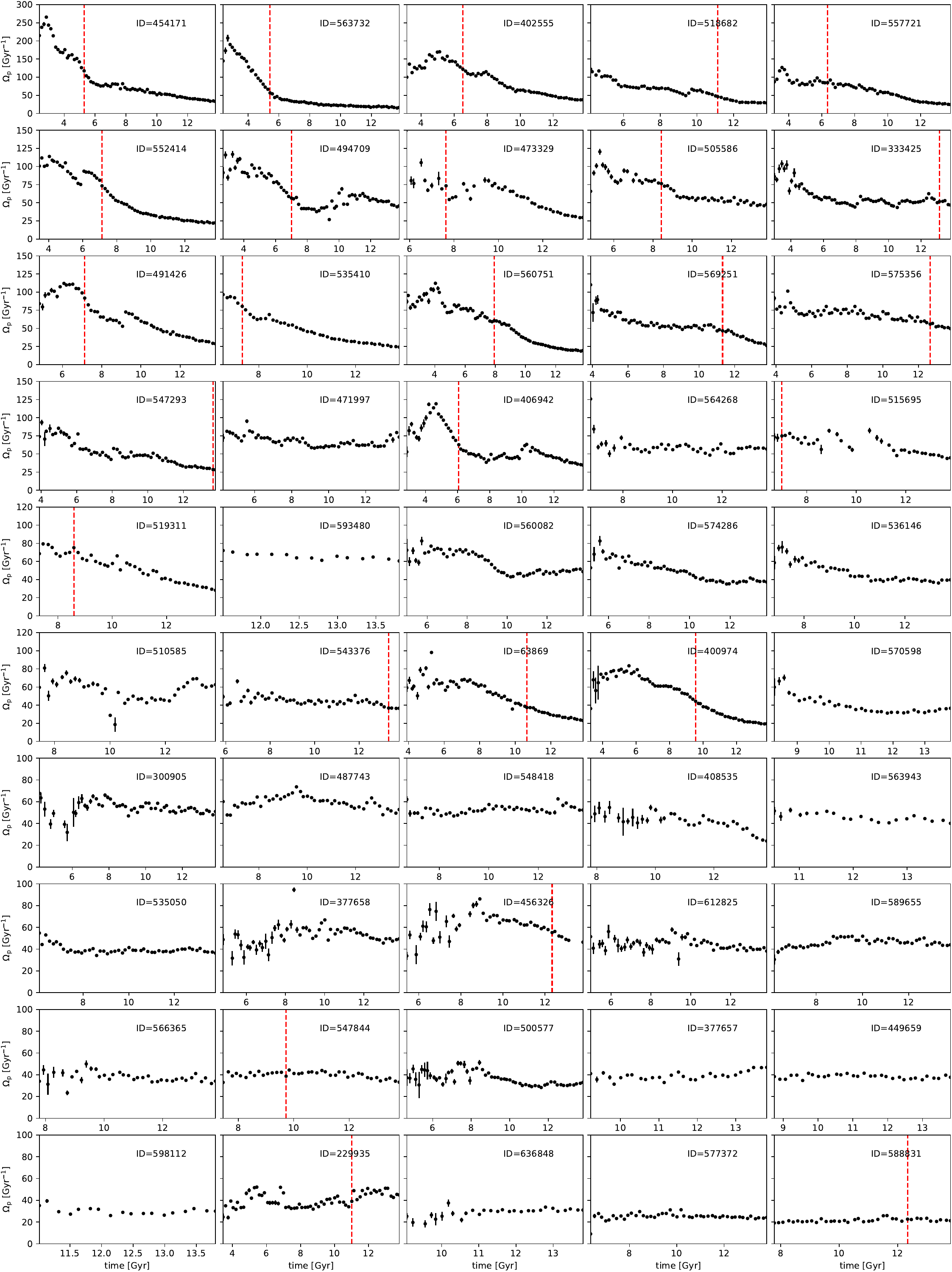}
\caption{Time evolution of the pattern speed for remaining cases from our sample of barred galaxies from TNG50-1. Dashed vertical red lines indicate $t_\mathrm{AGN}$, as in Fig.~\ref{omegas} and ig.~\ref{jumps}.}
\label{rest}%
\end{figure*}

Fig.~\ref{rest} shows the time evolution of pattern speed of the remaining cases from our sample of simulated bars. It demonstrates the spectrum of behaviours, together with the frequency of fluctuations. 

\section{Bar slowdown and properties of galaxies}
Previous studies of the slowdown of bars often used idealised numerical experiments where the total amount of dark matter and baryons is conserved or some components are injected artificially to understand its sole influence. While these are instructive, they are also unrealistic, since galaxies accrete mass, form stars, and interact with other galaxies throughout their history. The evolution is typically strongest in the earlier phases, which often coincide with the bar formation. Cosmological simulations offer therefore a more realistic view on the combination of these effects on the bar evolution, albeit at a somewhat reduced numerical resolution.

\subsection{Absolute slowdown}
\begin{figure}
\centering
    \includegraphics[width=.99\columnwidth]{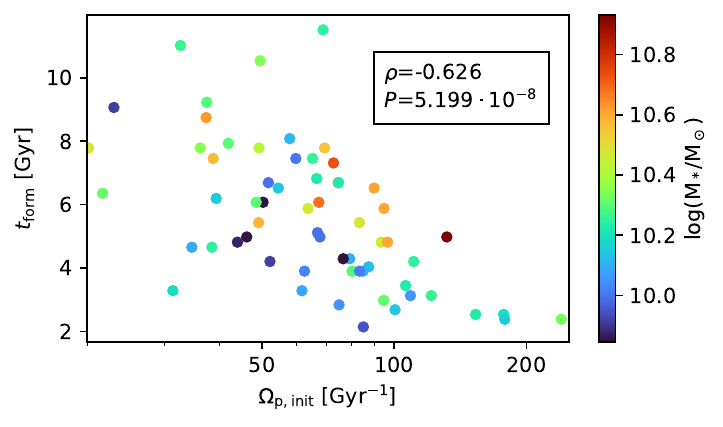}
    \caption{Correlation between the initial bar pattern speed $\Omega_{\mathrm{p, init}}$ and bar formation time $t_\mathrm{form}$. Colour indicates the total stellar masses of the simulated galaxies, averaged over the first six snapshots after bar formation. Spearman's correlation coefficient $\rho$ and its $p$-value are indicated.}
    \label{range}%
\end{figure}

From a quick look at the examples in Fig.~\ref{omegas} a simple conclusion can be drawn that simulated bars that form with absolute high values of $\Omega_\mathrm{p}$ slow down by a lot over their evolution, while those that start from lower values, retain a similar speed over its course. To make sure that this conclusion is true, a correct determination of the bar formation time is needed, since, if this is mistakenly derived by some automatic method at later times in cases from the top row of Fig.~\ref{omegas} it may appear similar to the cases for the bottom row. For this reason, we have determined the bar formation time by visual inspection of the surface density of stars in three projections (to make sure that the existence or not of the bar is not affected by the wrong inclination), a map of the average radial velocity $v_R$ and the radial profiles of $m=2$ Fourier analysis. In several cases, bars formed at high redshifts were destroyed and later a new bar emerged. When this happened we took the last bar formation time, which lasted until $z=0$ to consider in our analysis. In Fig.~\ref{range} we show the dependence of the initial pattern speed of bars $\Omega_{\mathrm{p, init}}$ (measured by averaging the values of the first six snapshots after bar formation in order to reduce noise) on the bar formation time $t_\mathrm{form}$ defined as above, colour coded according to the total stellar mass (also averaged in the first six snapshots after bar formation). Bars that start with larger values of $\Omega_{\mathrm{p}}$ tend to form earlier. The figure also shows that our sample covers a continuous range of $\Omega_{\mathrm{p, init}}$ and $t_\mathrm{form}$ values.     

\begin{figure}
\centering
    \includegraphics[width=.95\columnwidth]{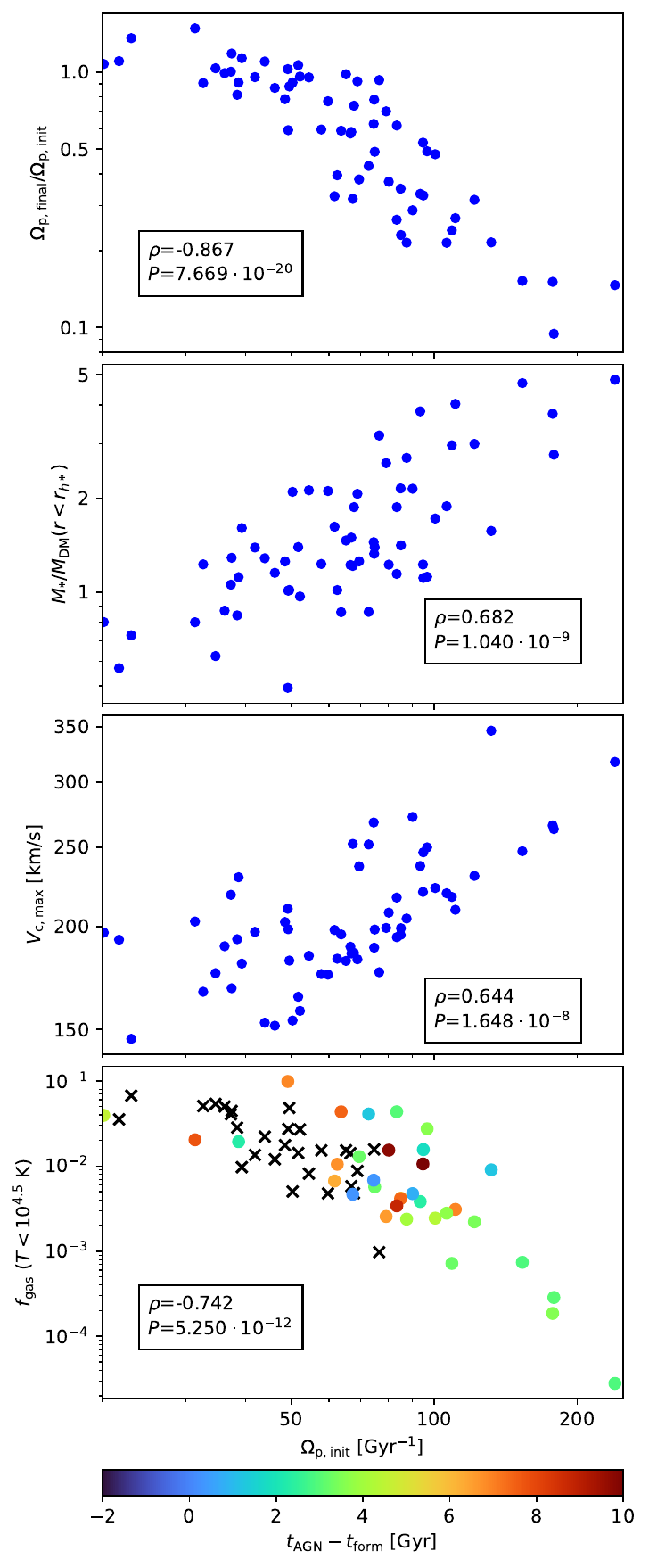}
    \caption{Correlations between the initial bar pattern speeds of $\Omega_{\mathrm{p, init}}$ and its ratio with the final value $\Omega_{\mathrm{p, final}}$ (top panel), the stellar-to-dark mass ratio measured within the stellar half-mass radius $M_{*}/M_{\mathrm{DM}} (r<r_{\mathrm{h}*})$ (second top panel), the maximum circular velocity $V_{\mathrm{c, max}}$ (second bottom panel), and the cold gas fraction $f_\mathrm{gas}$ ($T<10^{4.5}$ K) also measured inside $r<r_{\mathrm{h}*}$ (bottom panel). The values are averaged from six consecutive snapshots at the bar formation. The colour in the bottom panel marks the difference between the times when AGN feedback prescription switched from thermal to the kinetic and the bar formation times (or black crosses if this switch never occurred). In each panel, Spearman's correlation coefficients $\rho$ and its $P$-values are indicated.}
    \label{Corr_init}%
\end{figure}

In the top of Fig.~\ref{Corr_init}, we plot the initial bar pattern speeds against the final to initial ratios of the pattern speeds. We stress that the final and initial values used for this and the following plots
$X_\mathrm{final,initial}=\langle X \rangle_{N_\mathrm{avg}}$
were averaged in the first (or last) $N_\mathrm{avg}=6$ snapshots to smooth out any small deviations, such as those discernible in Figs.~\ref{omegas} and~\ref{jumps}. Changing the length of this averaging interval did not affect the general conclusions drawn from this and other plots. 
As inferred earlier from the examples in Figs.~\ref{omegas} and~\ref{jumps}, $\Omega_{\mathrm{p, init}}$ strongly determines the amount of slowdown of bars. This can be understood based on findings of \cite{Athanassoula2003}. Bars that have high values of $\Omega_{\mathrm{p}}$ have their resonances embedded at smaller angular momentum/energy in dark matter haloes, where density is higher and so more material is present that can take the angular momentum. While the statistics of this correlation can be increased by correlating $A$ vs $B/A$, this way of quantifying it does not determine the slope of this correlation. If there had been no slowdown, it would have been a straight line at $\Omega_{\mathrm{p, final}}/\Omega_{\mathrm{p, init}}\sim 0$.

Having established that $\Omega_{\mathrm{p, init}}$ is a crucial parameter in the evolution of $\Omega_{\mathrm{p}}$ we looked for additional correlations between $\Omega_{\mathrm{p, init}}$ and various galaxy properties. These were the masses of stars, gas (`cold', i.e. $T<10^{4.5}$\,K, star-forming and total), and dark matter measured within stellar half mass radius $r_{\mathrm{h}*}$, the ratio of stars to dark matter, the gas fractions, the half mass radii of stars and dark matter, $r_{\mathrm{h}*}$ and $r_{\mathrm{h,DM}}$, the initial values of $\mathcal{R}$ and the maximum circular velocities $V_{\mathrm{c, max}}$. To find the rotation curve and its maximum we calculated the accelerations of stellar particles using the \texttt{Griffin} code that employs the fast multipole method as force solver \citep{Dehnen2000, Dehnen2014}. 

We looked for correlations of the initial pattern $\Omega_{\mathrm{p, init}}$ with various global properties of the simulated galaxies. In the bottom three panels of Fig.~\ref{Corr_init}, we show those with Spearman correlation coefficient $|\rho|>0.6$. 
We found that $\Omega_{\mathrm{p, init}}$ correlates with stellar-to-dark mass ratios $M_{*}/M_{\mathrm{DM}}$ and the maximum circular speed $V_{\mathrm{c, max}}$, but anti-correlates with the cold gas fraction. We think that $M_{*}/M_{\mathrm{DM}}$ and cold gas fractions are affected by an artificial correlation between bar slowdown and AGN feedback (expanded more in the following section). In the bottom panel (plotting $f_\mathrm{gas}$ vs. $\Omega_{\mathrm{p, init}}$), we use colour to indicate the difference between $t_\mathrm{AGN}$ which coincides with the kinetic feedback driven central gas holes (defined in section 3) and the bar formation time. Galaxies that reached the kinetic feedback threshold at $t_\mathrm{AGN}$ are also more likely to have their gas heated by the AGN during the thermal feedback mode (since thermal energy emitted is proportional to the accretion rate, which at the same increases SMBH mass, which is a determining parameter of the switch between the two modes), which will reduce the amount of the cold gas.   

The correlation between $V_{\mathrm{c, max}}$ and $\Omega_{\mathrm{p, init}}$ can have two explanations. First, in more massive haloes everything rotates faster and this might transfer to newly formed bars. Another explanation is that, since the particle resolution is fixed in this simulation, more massive galaxies are better resolved and therefore can experience this initial phase of very high absolute pattern speed, while for the less massive objects, low resolution prevents this from happening.   

\subsection{Relative slowdown}
As shown in the previous subsection the initial absolute pattern speed and therefore the absolute slowdown of bars scales with $V_{\mathrm{c, max}}$ of galaxies. This scaling means that comparing the absolute values may be a bit misleading. To look at the problem in a more dimensionless manner, following \cite{Chiba2021} we adopt the slowing rate parameter $\eta=-\dot{\Omega}_{\mathrm{p} }/\Omega_\mathrm{p}^2=\mathrm{d}\Omega_{\mathrm{p}}^{-1}/\mathrm{d}t$ and define an averaged slowing rate more suited for our numerical study
\begin{equation}
    \langle \eta \rangle =-\frac{\Delta \Omega_{\mathrm{p}}/ \Delta t}{\Omega_{\mathrm{p,init}}\,\Omega_{\mathrm{p,final}}}
    = \frac1{\Delta t}
    \left[\frac1{\Omega_{\mathrm{p,final}}}-\frac1{\Omega_{\mathrm{p,init}}}\right],
\end{equation}
where $\Delta t$ is the time between measuring $\Omega_{\mathrm{p,final}}$ and $\Omega_{\mathrm{p,initial}}$.

\begin{figure}
\centering
\includegraphics[width=.95\columnwidth]{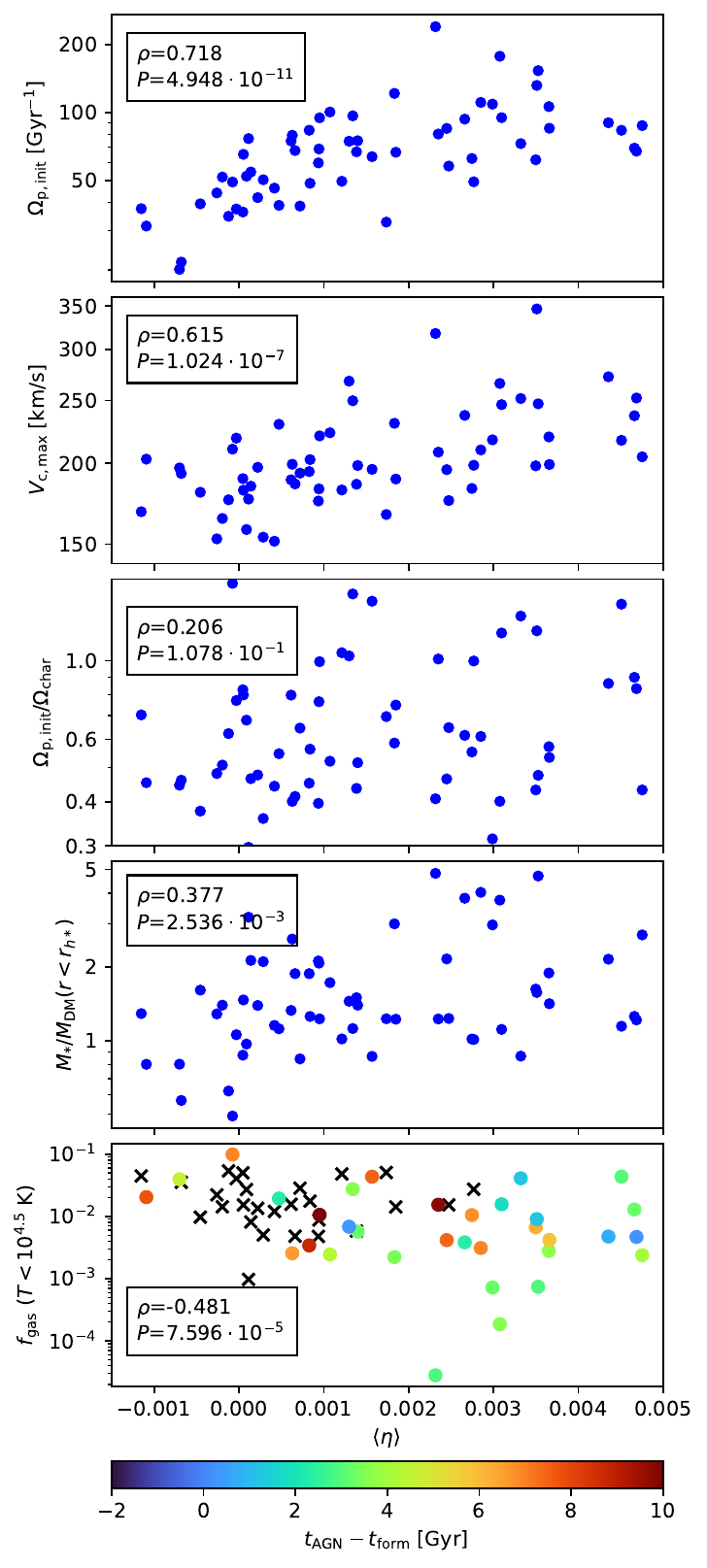}
\caption{Correlations between the mean slowdown parameter $\langle\eta\rangle$ and the initial pattern speeds of bars $\Omega_{\mathrm{p, init}}$ (top), the maximum circular velocity $V_{\mathrm{c, max}}$ (second top), normalised initial pattern speeds of bars $\Omega_{\mathrm{p, init}}/\Omega_\mathrm{char}$ (middle), the stellar-to-dark mass ratio measured within the stellar half-mass radius $M_{*}/M_{\mathrm{DM}} (r<r_{\mathrm{h}*})$ (second bottom), and the cold gas fraction $f_\mathrm{gas}$ ($T<10^{4.5}$ K) also measured inside $r<r_{\mathrm{h}*}$ (bottom). The colour scale for the bottom panel is the same as for Fig.~\ref{Corr_init}.
}
          \label{eta}%
\end{figure}

The two top panels of Fig.~\ref{eta} show the correlations between $\langle \eta \rangle$ and $\Omega_{\mathrm{p,init}}$ and $V_{\mathrm{c, max}}$. Mean slowing rate scales with these two values, just like $\Omega_{\mathrm{p, final}}/\Omega_{\mathrm{p, init}}$. However, when we normalize the $\Omega_{\mathrm{p, init}}$ with a characteristic frequency defined as $\Omega_{\mathrm{char}}=\Omega_{\mathrm{c}} (r_{\mathrm{h}*})$ the correlation is no longer there (middle panel of Fig.~\ref{eta}). This is unsurprising, since $\Omega_{\mathrm{char}}$ approximately scales with $V_\mathrm{c,max}$, which as we found correlates with $\Omega_{\mathrm{p, init}}$. The lack of correlation in this space reinforces that the galaxy mass (and therefore resolution) scaling may be the dominant factor determining $\Omega_{\mathrm{p, init}}$. The weakening of the correlations is also seen in the two bottom panels of Fig.~\ref{eta} where we look into the initial stellar to dark mass ratio and cold gas fractions (already dimensionless).     

\subsection{Slowdown and the evolution of bars}
\begin{figure}
\centering
\includegraphics[width=.95\columnwidth]{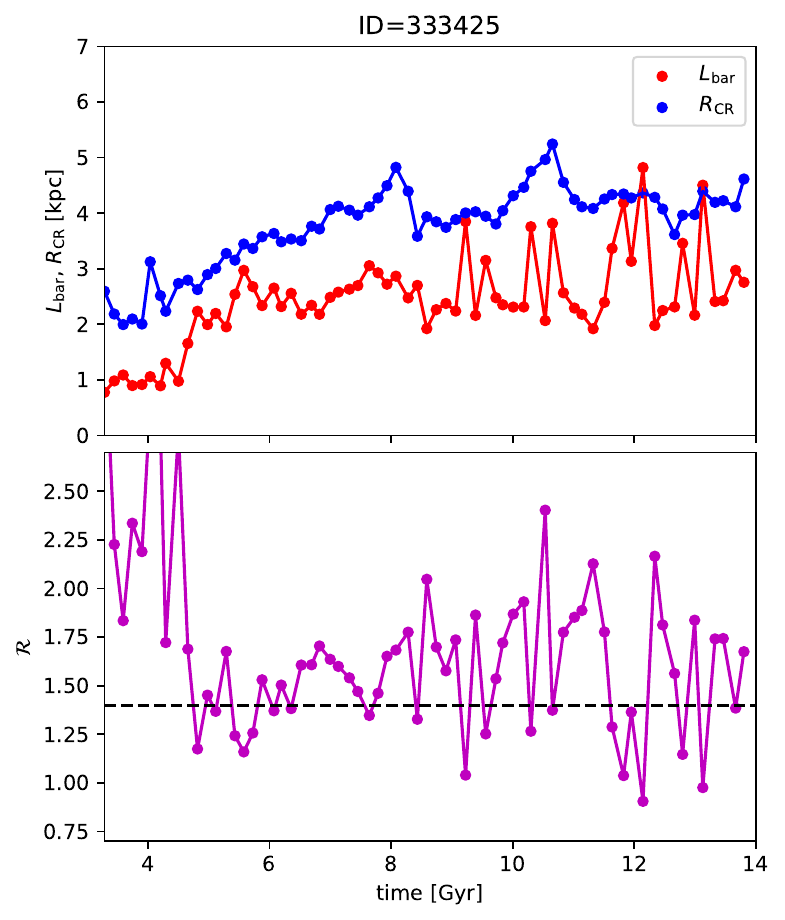}
\caption{Top: An example of the time evolution of the bar length $L_{\mathrm{bar}}$ and the corotation radius $R_{\mathrm{CR}}$ for a barred galaxy of ID=333425 at $z=0$. Bottom: The time evolution of the ratio $\mathcal{R}=R_{\mathrm{CR}}/L_{\mathrm{bar}}$ of the same galaxy as in top. A black dashed horizontal line marks the customary boundary of $\mathcal{R}=1.4$ between the fast and slow bars. While the average evolution of the bar length and the corotation radius is stable, small deviations in both of them make the ratio noisy which complicates the interpretation of the values of this parameter.}
          \label{bad_R}%
\end{figure}

When bars slow down, the position of the bar resonances moves outwards, thus enabling the trapping of new stars and an increase in bar size. It is still under debate at what rate both things occur, which can result in the evolution of $\mathcal{R}$. It is long known that $\mathcal{R}$ is a problematic parameter to calculate both in observations and simulations. Because it is a ratio, a small deviation in the denominator can change significantly its value and physical interpretation. It was shown in simulations several times (e.g. \citealt{MD2006}; \citealt{Ghosh2024}) that the bar length, i.e. the denominator in $\mathcal{R}$ is a very problematic parameter to measure. It is also true for the methods we used for the analysis of this simulation and an example is shown in Fig.~\ref{bad_R}. In the top panel of this Figure, we show the time evolution of the bar length $L_\mathrm{bar}$ assumed to be the outer radius of the bar region, as calculated by \texttt{patternSpeed.py} \citep{Dehnen2023} and the corotation radius $R_{\mathrm{CR}}$, which was calculated using the pattern speed measurements and the rotation curves as described in 4.1. We note that the primary purpose of the outer bar region given by \texttt{patternSpeed.py} is to estimate $\Omega_\mathrm{p}$ and using it as a $L_\mathrm{bar}$ is a rough proxy. The bottom panel shows the time evolution of the ratio of these two parameters, i.e. $\mathcal{R}$ with the marked boundary separating the slow and fast bars. We see that while the average time evolution of the $L_\mathrm{bar}$ and $R_{\mathrm{CR}}$ shows a stable increasing trend, its ratio has a very noisy pattern, where the classification from the slow to fast bar changes multiple times. $R_{\mathrm{CR}}$ fluctuates in this example galaxy, because of its strong spiral structure, as described by \cite{Wu2016}. Because of the difficulties presented in this example, we abstain from analysing the time evolution of $\mathcal{R}$ as the noise may lead to wrong conclusions, and instead focus on looking directly at the co-evolution of bar slowdown and growth of the bar.
 
\begin{figure}
\centering
\includegraphics[width=.95\columnwidth]{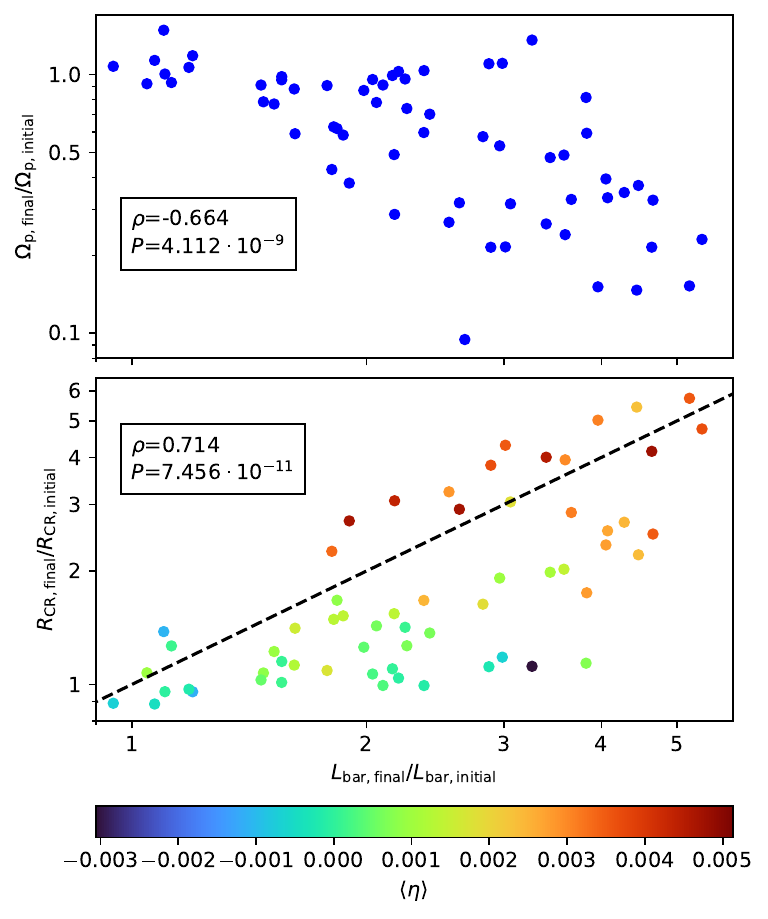}
\caption{Correlations between the final to initial ratios of pattern speeds $\Omega_\mathrm{p}$ and bar lengths $L_\mathrm{bar}$ (top) and corotation radii $R_\mathrm{CR}$ (bottom).
The dashed black line in the bottom panel marks where $\mathcal{R}$ would not change, i.e. $\mathcal{R}_\mathrm{final}/\mathcal{R}_\mathrm{init}=1$. The colour of points in the top panel indicates the mean slowing rate $\langle \eta \rangle$. The bars that slow down more lie above the $\mathcal{R}_\mathrm{final}/\mathcal{R}_\mathrm{init}=1$ which means their $\mathcal{R}$ goes up and they slow down also in terms of $\mathcal{R}$.}
          \label{Corr_Lbar}%
\end{figure}

 Fig.~\ref{Corr_Lbar} shows the correlations between the ratios of the final to initial values of bar lengths $L_\mathrm{bar}$ 
 and pattern speeds $\Omega_\mathrm{p}$ (top) and corotation radii $R_\mathrm{CR}$ (bottom). Both figures confirm that when bars slow and the corotation radii grow, they also increase in size. The growth of the corotation radius is greater than the growth of the bar size for the galaxies that slow down more and reside in more massive galaxies. In the bottom panel of Fig.~\ref{Corr_Lbar}, we plot a line of $\mathcal{R}_\mathrm{final}/\mathcal{R}_\mathrm{init}=1$ and colour the points with values of the mean slowing rate $\langle \eta \rangle$. Bars that slow down more lie over the plotted line, therefore their $\mathcal{R}$ would increase and they would slow down also in terms of this parameter. Bars with low values of $\langle \eta \rangle$ lay below the dashed line, which would imply that they speed up in terms of $\mathcal{R}$. While their $\Omega_\mathrm{p}$ remains on average constant throughout their evolution, the decrease in $\mathcal{R}$ is caused by the noise, as in the case of the example in Fig.~\ref{bad_R}. While the bar length is already a noisy parameter, its initial values suffer more from noise, since bars are usually weak in their early evolutionary stages and strengthen with time.     

\begin{figure}
\centering
\includegraphics[width=.95\columnwidth]{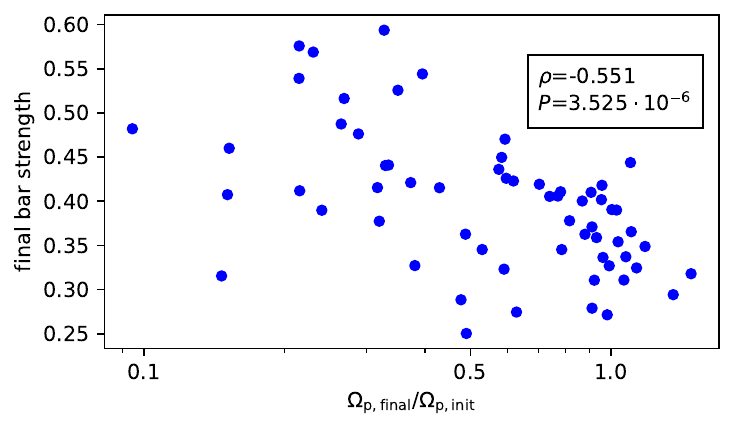}
\caption{Correlation between the final to initial pattern speed ratio and the final bar strength, measured as the maximum of the $m=2$ Fourier amplitude. Bars that have slowed down significantly tend to grow stronger.}
          \label{bar_str}%
\end{figure}

Besides looking at the growth of the bars in their sizes, we also looked at how their final strength is related to the slowdown. Fig.~\ref{bar_str} shows the correlation between $\Omega_{\mathrm{p, final}}/\Omega_{\mathrm{p, init}}$ and the final bar strength. The bar strength here is defined as the maximum of the $m=2$ Fourier amplitude. We find that bars that slowed down significantly tend to end up stronger than those which retain their pattern speed approximately constant. Bars that do not slow down do not reach amplitudes higher than 0.4-0.5. This correlation is in agreement with findings of \cite{Athanassoula2003} from the isolated simulations.

\section{Bar slowdown and AGN feedback relation}
\label{sec:AGN}
\begin{figure}
\centering
\includegraphics[width=.99\columnwidth]{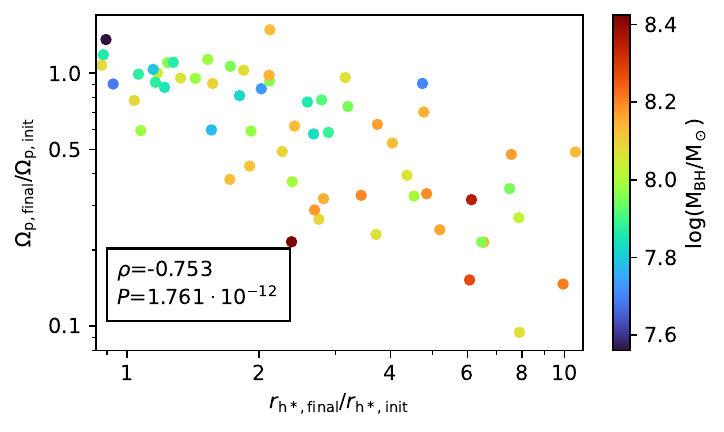}
\caption{ 
    The correlation between the final to initial ratios of the stellar half-mass radii of the barred galaxies and the bar pattern speeds. The colour bar marks the final mass of the SMBH of those galaxies. Those galaxies which SMBH lie above the threshold of $10^8\,$M$_\odot$ had their AGN feedback switch to kinetic mode. We argue in the text that this correlation is at least partly a result of the AGN feedback prescription of IllustrisTNG.
}\label{BH}%
\end{figure}

\begin{figure*}
\centering
\includegraphics[width=18.5cm]{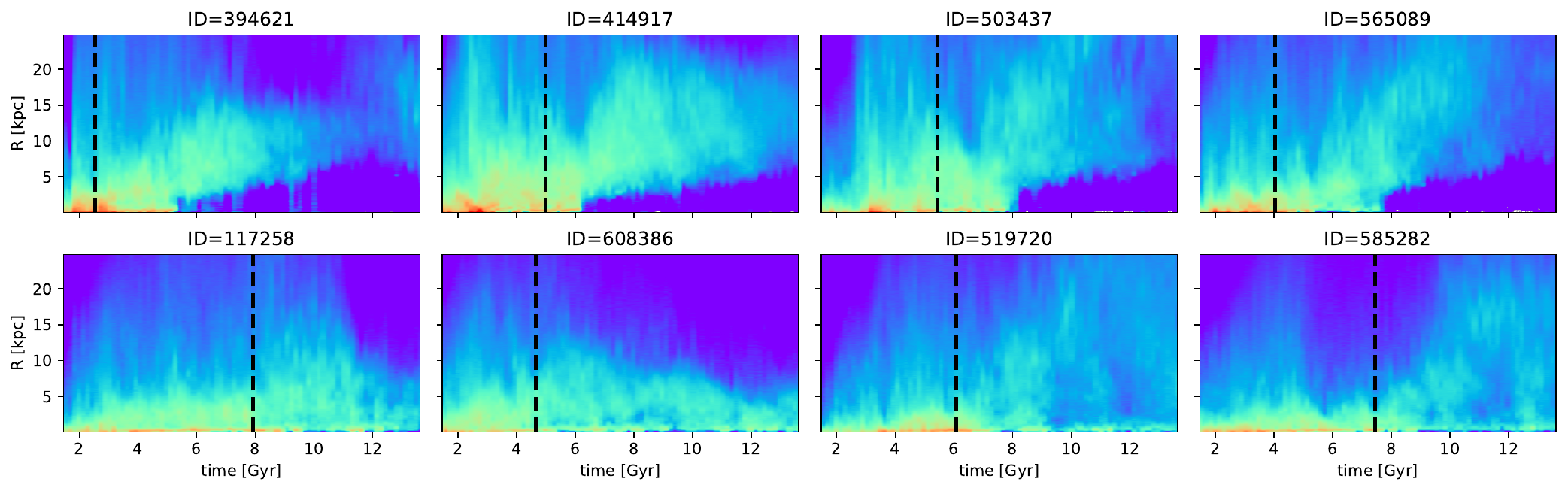}
\caption{Time evolution of the gas density profiles for eight example simulated barred galaxies from TNG50-1. The IDs given in the plots are for $z=0$. Vertical black dashed lines mark the bar formation times. The examples correspond to the cases from Fig.~\ref{omegas} where those in the top panel experienced a significant slowdown and those in the bottom panel had smaller changes in $\Omega_{\mathrm{p}}$.}
\label{gas}%
\end{figure*}

The AGN feedback in the IllustrisTNG suite is constructed with two modes: thermal and kinetic \citep{Weinberger2017, Weinberger2018}. The SMBH is first seeded with an initial mass of $1.2\times10^{6}\;\mathrm{M}_\odot$, when the mass of the galaxy exceeds $7.4\times10^{10}\,\mathrm{M}_\odot$, and thereafter grows through mergers and gas accretion. AGN feedback is modelled initially in a thermal model, where energy is transferred to the surrounding gas. Once the SMBH reaches a mass of $10^{8}\;\mathrm{M}_\odot$, the feedback prescription is switched from thermal to kinetic mode, where both energy and momentum are transferred to the gas. Shortly afterwards, at around time $t_{\mathrm{AGN}}$, the feedback quickly creates a large central hole in the gas distribution, reaching 2-8\,kpc in radius.

We noticed a correlation between the time evolutions of bar pattern speeds and stellar half-mass radii: cases of decreasing $\Omega_{\mathrm{p}}$ have growing $r_{\mathrm{h}*}$, while for near-constant $\Omega_{\mathrm{p}}$ also $r_{\mathrm{h}*}$ stayed constant. This behaviour is summarized in Fig.~\ref{BH}. The explanation is that bars that slow down more reside in galaxies that have their SMBHs grow earlier and faster. For these galaxies, the simulation prescriptions switch feedback earlier to kinetic mode which clears the gas from the inner parts of the galaxy. 
The formation of these holes and the distribution of gas, which is a star formation fuel can be seen in the top row of Fig.~\ref{gas} where the examples are the same as in Fig.~\ref{omegas}.
This gas removal has two effects. First, it limits star formation to the outer parts, which results in enhanced increase of $r_{\mathrm{h}*}$. Second, removing cold gas from the vicinity of the bar prevents gas inflow and the associated angular momentum gain by and weakening of the bar, i.e.\ shifts the balance towards bar slowdown and strengthening. Thus we expect the correlation between $r_{\mathrm{h}*}$ and the amount of slowdown of simulated bars is created or at least exacerbated by the specific way AGN feedback is implemented in IllustrisTNG.

Another consequence of this relation is that average galaxy properties measured at or within $r_{\mathrm{h*}}$ are affected by the AGN-feedback driven evolution of $r_{\mathrm{h*}}$, confusing the interpretation of any correlations.

\section{Discussion and summary}
\subsection{Discussion}
We found a continuous distribution of pattern speed evolutions in the considered sample. In terms of the evolution of the absolute values, the behaviour ranges from those that start at very high pattern speed, $\Omega_\mathrm{p}$, and later slow down drastically, to those for which $\Omega_\mathrm{p}$ start at lower values and remain more or less constant. We found that the change in this behaviour correlates with the galaxy mass (quantified by $V_\mathrm{c, max}$). Since the stellar particle masses in TNG simulations do not vary much, the more massive galaxies are better resolved than the less massive ones. This suggests that evolution differences in $\Omega_\mathrm{p}$ in this simulation suite are affected by resolution. Indeed, \cite{Frankel2022} found that $\Omega_\mathrm{p}$ of bars simulated at 8 times lower mass resolution (TNG50-2 vs. TNG50-1 with otherwise identical initial conditions) is on average more than twice smaller. Similarly, the (initial) $\Omega_\mathrm{p}$ of bars formed in isolated idealized simulations without gas decreases with resolution \citep[e.g.][]{SD2009}, though much less so than found by \citeauthor{Frankel2022}.

This resolution effect exacerbates the effect shown in Fig.~\ref{BH} and discussed in section~\ref{sec:AGN} of enhanced bar slowdown as consequence of gas removal due to stronger AGN feedback in more massive galaxies. As this latter is at least to some degree unrealistic, it appears that simulated bar slowdown is plagued by numerical artefacts in several ways.

The AGN feedback prescription is another factor that is tied to galaxy mass (and resolution) and shapes the bar pattern speed evolution, as discussed in section~\ref{sec:AGN}. More massive galaxies have the AGN switch to kinetic feedback earlier, removing the cold gas from their inner parts and preventing gas inflow which would otherwise oppose the slowdown. Therefore, it is not yet clear to which degree the mechanism of the bar gaining angular-momentum from the gas and losing to the halo, which was confirmed in idealised isolated simulations \citep{Friedli1993, VV2009, VV2010, Athanassoula2013, Athanassoula2014, Beane2023}, applies to more realistic simulations or even real galaxies.

A natural next step in the investigation of physics behind bar slowdown in cosmological setups, with ever-changing galaxy components, is to use higher resolution, but also a range of ideally more realistic sub-grid prescriptions of all unresolved processes, but in particular AGN feedback. This would show to which degree the simulations converge and to which degree the results reported here are artefacts of the simulation prescriptions or are general laws ruling the bar pattern speed evolution. While higher-resolution cosmological runs are very expensive, zoom-in simulations like Auriga \citep{Grand2016} are more affordable and can give some partial answers. The changes in terms of $\mathcal{R}$ \citep{Fragkoudi2021} suggest that the evolution of $\Omega_\mathrm{p}$ is also different from TNG50.

\subsection{Summary}
In this paper, we have studied the pattern speed evolution of a sample of 62 simulated barred galaxies from the TNG50-1 run of the magnetohydrodynamical cosmological suite IllustrisTNG \citep{Pillepich2019, Nelson2019b, Nelson2019}.
We used our recent code \citep{Dehnen2023} to measure the bar pattern speeds from individual snapshots. We found bars that start their evolution with high absolute values of $\Omega_\mathrm{p}$ and quickly slow down. We also found bars whose $\Omega_{\mathrm{p}}$ start at lower values and remain approximately constant until the end of the simulation.
The transition between the two behaviours seems to be smooth without any apparent gaps. We find the former behaviour more frequently in more massive and thus more resolved galaxies, while the latter happens more often in the low-resolution cases.
The prescription for AGN feedback in the TNG simulations results in the more massive galaxies having more intense feedback, which leads to a coincidence between the bar slowdown and the removal of the cold gas from the inner parts of the disc. Most of the simulated bars which did slow down are found in galactic discs with inner holes in their gas discs. Had this gas not been removed, it could potentially prevent the slowdown, according to previous experiments on isolated galaxies.

A central point arising from our analysis is the connection between the specific AGN feedback model and bar slowdown. This differs from commonly made arguments which target how bars feed the AGN, or how bars drive quenching. Here, we have outlined how excavation of the inner galaxy ISM by AGN feedback drives accelerated bar slowdown, which strongly depends on how long and to which extent the cold ISM persists. This dynamical behaviour thus provides an independent way to distinguish AGN-driven quenching by removing the star-forming gas or by starving it via removal of the coronal gas from which the star-forming gas is replenished.

\begin{acknowledgements}
 We are grateful to the IllustrisTNG team for making their simulations publicly available. We appreciate insightful discussions with T. Antoja, E. L. \L{}okas, V. P. Debattista, M. Bernet, P. Ramos, J. Ardèvol, F. Figueras, F. Anders, M. Romero Gómez. This work was partially supported by the Spanish MICIN/AEI/10.13039/501100011033 and by "ERDF A way of making Europe" by the “European Union” and the European Union «Next Generation EU»/PRTR, through grants PID2021-125451NA-I00 and CNS2022-135232, and the Institute of Cosmos Sciences University of Barcelona (ICCUB, Unidad de Excelencia ’Mar\'{\i}a de Maeztu’) through grant CEX2019-000918-M. RS gratefully acknowledges the generous support of a Royal Society University Research Fellowship. This work was performed using the DiRAC Data Intensive service at Leicester, operated by the University of Leicester IT Services, which forms part of the STFC DiRAC HPC Facility (\url{www.dirac.ac.uk}). The equipment was funded by BEIS capital funding via STFC capital grants ST/K000373/1 and ST/R002363/1 and STFC DiRAC Operations grant ST/R001014/1. DiRAC is part of the National e-Infrastructure.\\

 This work made use of the following software packages: \texttt{matplotlib} \citep{Hunter:2007}, \texttt{numpy} \citep{numpy}, \texttt{python} \citep{python}, \texttt{scipy} \citep{2020SciPy-NMeth, scipy_11255513}, \texttt{Cython} \citep{cython:2011}, \texttt{h5py} \citep{collette_python_hdf5_2014, h5py_7560547} and \texttt{Numba} \citep{numba:2015, Numba_11642058}.

Software citation information aggregated using \texttt{{https://www.tomwagg.com/software-citation-station/}{The Software Citation Station}} \citep{software-citation-station-paper, software-citation-station-zenodo}.
\end{acknowledgements}

\bibliographystyle{mnras}
\bibliography{bars}

\end{document}